\documentclass[a4paper,american]{scrartcl}
\usepackage{lmodern}
\usepackage[utf8]{inputenc}
\usepackage[T1]{fontenc}
\usepackage{color}
\usepackage{todonotes}
\usepackage{babel}
\usepackage{amsmath}
\usepackage{amssymb}
\usepackage{setspace}
\usepackage[authoryear]{natbib}
\usepackage[font=small]{quoting}
\usepackage{enumerate}
\usepackage{latexsym}
\usepackage[bookmarks=false,pdfborder={0 0 0},colorlinks=false]{hyperref}

\makeatletter

\renewcommand{\Im}{\operatorname{Im}}
\newcommand{\IR}{\mathbb{R}}

\makeatother

\begin{document}

\title{From the universe to subsystems: Why quantum mechanics appears more stochastic than classical mechanics}

\author{Andrea Oldofredi%
\thanks{Université de Lausanne, Faculté des lettres, Section de philosophie,
1015 Lausanne, Switzerland. E-mail:
\protect\href{mailto:Andrea.Oldoffredi@unil.ch}{Andrea Oldofredi@unil.ch}}%
, Dustin Lazarovici%
\thanks{Université de Lausanne, Faculté des lettres, Section de philosophie,
1015 Lausanne, Switzerland. E-mail:
\protect\href{mailto:Dustin.Lazarovici@unil.ch}{Dustin.Lazarovici@unil.ch}}%
, Dirk-André Deckert%
\thanks{Ludwig-Maximilians-Universität München, Mathematisches Institut,
Theresienstrasse 39, 80333 München, Germany. E-mail: \protect\href{mailto:deckert@math.lmu.de}{deckert@math.lmu.de}%
}, Michael Esfeld%
\thanks{Université de Lausanne, Faculté des lettres, Section de philosophie,
1015 Lausanne, Switzerland. E-mail: \protect\href{mailto:Michael-Andreas.Esfeld@unil.ch}{Michael-Andreas.Esfeld@unil.ch}}
}

\maketitle
\begin{abstract}
 \begin{center} Forthcoming in Fluctuations and Noise Letters, Special issue Quantum and classical frontiers of noise    \end{center}
    \medskip

By means of the examples of classical and Bohmian quantum mechanics, we illustrate the well-known ideas of Boltzmann as to how one gets from laws defined for the universe as a whole to dynamical relations describing the evolution of subsystems. We explain how probabilities enter into this process, what quantum and classical probabilities have in common and where exactly their difference lies.  
\medskip{}

\end{abstract}

\tableofcontents{}

\section{Introduction}
\label{sec:introduction}

Any fundamental physical theory is a theory of the universe as a whole: its laws describe the evolution of the \emph{entire} configuration of matter. Thus, in classical mechanics (CM), where the forces range all over physical space, the motion of any particle at any given time depends, strictly speaking, on the position and the momentum of all the other particles and thus on the initial state of the entire universe. In quantum mechanics (QM), due to entanglement, the only fundamental quantum state is the one pertaining to the universe as a whole and represented by the universal wave function. However, this fundamental point of view is utterly impractical for everyday science, which seeks to apply these theories to quite small parts of the universe. Aside from our limited computational resources, we simply do not know the exact configuration of matter and / or the exact wave function of the universe so that we could solve the equations of motion for them. 

Thus, in order to derive testable propositions from a physical theory, we need a procedure to get from fundamental laws, describing the global evolution of the universe, to predictions about particular subsystems. Such a procedure was proposed by Ludwig Boltzmann whose derivation of thermodynamic laws from microscopic particle dynamics can be viewed as a general scheme for probabilistic reasoning in the face of incomplete information.

The aim of this paper is to illustrate Boltzmann's ideas as a general way of understanding probabilities in deterministic theories. We argue that the same reasoning applies to both classical and quantum mechanics, in particular if the latter is understood in terms of Bohmian mechanics. In fact, as Einstein already noted, Boltzmann's insights are independent of the details of the underlying microscopic theory (c.f. Einstein's \emph{autobiographical notes} in \cite{schilpp}). Against this background, we then inquire what, if any, is the difference between probabilities in CM and in QM and why the quantum world appears to us so much more random and unpredictable.

Concerning quantum mechanics, we endorse the theory going back to \cite{Broglie:1928aa} and \cite{Bohm:1952aa} whose dominant contemporary version is known as Bohmian Mechanics (BM) (\cite{Durr:2013aa}). The primary reason for doing so is that QM runs into the infamous measurement problem illustrated by Schrödinger's cat paradox (see \cite{Maudlin:1995aa} for a precise formulation). Quantum theories that solve the measurement problem by being committed to a definite configuration of matter in physical space are known as primitive ontology theories. The wave function then has the job to describe how this configuration evolves in time. Bohmian mechanics is the most prominent example of a primitive ontology formulation of QM. The primitive ontology here are particles characterized by their positions. The configuration of particles in physical space then evolves according to a non-local law of motion in which the wave function enters. We adopt Bohmian mechanics, because we take it to provide the most convincing solution to the measurement problem (but we do not have the space to argue for this claim here; see e.g. \cite{Maudlin:1995ab} and \cite{Esfeld:2014ac}).

Moreover, by positing the same primitive ontology as CM -- point particles moving in three-dimensional physical space -- BM is best suited for highlighting the similarities between quantum and classical mechanics as far as the status and interpretation of probabilities is concerned. In this vein, we seek to counter the widespread belief that probabilities in the quantum realm are fundamentally different from those encountered in classical statistical mechanics. (In standard QM, the source of randomness is the collapse of the wave function replacing the deterministic Schrödinger evolution. But since the collapse postulate is obscure, the status of this randomness remains obscure as well.)

Nonetheless, there are certain striking differences between CM and QM that we have to account for. For instance, also in Bohmian QM, one cannot do better than to make statistical predictions according to Born's rule. In CM, by contrast, there are many situations in which one can obtain a reliable deterministic description of a particular subsystem. What is the reason for this difference? We seek to answer these questions in the concluding Section 4 of the paper. 

\section{Probabilities in classical mechanics}
\label{sec:prob-classical}

In classical mechanics, the physical state of an $N$-particle system is completely determined by specifying the positions and momenta of all the particles. Denoting by $q_i$ and $p_i$ the position, respectively the momentum of the i'th particle, we call $X(t)=(q_1(t),...,q_N(t); p_1(t),...,p_N(t))$ the \emph{microstate} of the system at time $t$. The space of all possible microstates, here $\Gamma := \IR^{3N} \times \IR^{3N}$, is called \emph{phase space}. The microstate evolves according to the microscopic laws of motion, which, in the Hamiltonian formulation, take the form
\begin{equation}\label{Hamiltonianeq}
 \begin{cases}
\dot{q_i}
=
\; \frac{\partial H}{\partial p_i}
\\[1.5ex]
\dot{p_i}
=
-\frac{\partial H}{\partial q_i} \end{cases},
\end{equation}
\noindent  with 
\begin{equation}\label{Hamiltonian}
H(q,p)=\sum\limits_{i=1}^N\frac{p_i^2}{2m_i}+V(q_1, \dots, q_n). \end{equation}
More compactly, this can be written as
\begin{align}
\label{eq:Hamilton-eq}
(\dot{q_i},\dot{p_i})
=
v^{H}(q,p),
\end{align}

\noindent where $v^{H}$ denotes the vector field on $\Gamma$ generated by the Hamiltonian $H$. These equations give rise to a Hamiltonian flow $\Phi_{t,0}$ such that $X(t) = \Phi_{t,0}(X)$ for any initial microstate $X$. In equation \eqref{Hamiltonian}, $m_i$ denotes the mass of the i'th particle and $V$ the interaction potential, which can be split into

\begin{equation} V(q_1, \dots, q_n) = \sum\limits_{i<j} V_{int}(q_i - q_j) +  V_{ext} (q_1, ... , q_N, t). \end{equation}

\noindent $V_{int}$ then corresponds to a pair-interaction among the particles (e.g.  gravitation) and $V_{ext}$ is an external potential, summarizing the influences of the environment. Of course, if the $N$ particle system is the entire universe, then $V_{ext} = 0$, since there is nothing outside the universe.

If $V_{ext}$ is zero (or at least time-independent), such a Hamiltonian system has several nice properties. For one, it conserves the total energy, meaning that $H = const.$ along any solution of \eqref{Hamiltonianeq}. Furthermore, by the Liouville theorem, the Hamiltonian flow conserves phase space volume. This is to say that the uniform Lebesgue measure $\lambda$ is a \emph{stationary} measure on $\Gamma$ in the sense that for all $t \geq 0$ and any Borel set $A \subseteq \Gamma$,
\begin{equation}\label{stationarity} \lambda(\Phi_{t,0} A) = \lambda(A).\end{equation}

\noindent For fixed $E \in \IR$, it is usually convenient to consider the reduced phase space $\Gamma_{E} := \lbrace X \in \Gamma : H(X) = E \rbrace$ to which a system with total energy $E$ is confined by virtue of energy conservation. $\lambda$ then induces a stationary measure $\lambda_E$ on the hypersurface $\Gamma_E$, which is called the \emph{microcanonical measure}. By convention, we normalize this measure to $\lambda_E(\Gamma_E) = 1$.

\subsection{Randomness and typicality}

\noindent Given that CM is deterministic, where does randomness come from? There are at least three reasons to depart from the deterministic description: (i) we do not have access to the exact values of all positions and momenta in a given physical system. We can neither manipulate them with arbitrary precision in experimental situations, nor measure the exact (initial) microstate $X$  in order to determine the system's trajectory. (ii) Physical systems can be extremely sensitive to perturbations of their initial conditions. This means that even a small error about the initial data can translate into a huge error about the evolution of the system. (iii) The complexity of calculation increases rapidly as $N$ becomes very large.

Against this background, it seems reasonable and necessary to make two concessions. First, it usually suffices to provide a \emph{coarse-grained} description of the system. That is, rather than asking for the exact microstate, we are interested in the value of certain macroscopic ``observables'' $F: \Gamma \to \IR$. These observables are coarse-graining in the sense that a great number of microstates $X$ will in general correspond to (approximately) the same value of $F$.\footnote{Mathematically, if $\Gamma$ is endowed with a probability measure, such a function is called a \emph{random variable}, though the name is somewhat deceiving: there is really nothing random about it, since the \emph{macrostate} of a system (defined in terms of such observables) is determined by its microstate.} Second, since we cannot determine the exact evolution of the system -- if only for the fact that we do not know the exact initial conditions --, we can only ask what happens in \emph{most} possible instances, that is, for \emph{typical} initial conditions.

In some cases, typical trajectories coarse grain to one and the same macroscopic history, so that predictions appear deterministic (e.g. when we set out to determine the trajectory of a stone thrown on earth). In many cases, though, typical initial conditions agree only on certain statistical patterns in the distribution of coarse-grained observables (e.g. when we ask for the relative frequency of heads or tails in a long series of coin tosses). In these cases, probabilities come into play.

In any case, if we can establish that a certain fact or feature occurs for the vast majority of possible initial conditions -- that is, in the last resort, for the overwhelming majority of possible universes described by a particular theory --, we can justifiably call it a prediction of that theory (see \cite{Typicality} for a detailed discussion). In order to make such an argument precise, we need a measure on phase space telling us what an ``overwhelming majority'' of initial conditions is. Such a measure is called \emph{typicality} measure. Given a typicality measure $\mu$, we can say that a particular property $P$ is \emph{typical}, if
\begin{equation} \mu \Bigl( \Big\lbrace X \in \Gamma : P(X) \Bigr\rbrace \Bigr) \approx 1, \end{equation}

\noindent and \emph{atypical} if 

\begin{equation} \mu \Bigl( \Big\lbrace X \in \Gamma : P(X) \Bigr\rbrace \Bigr) \approx 0. \end{equation}
Note that most properties will be neither typical nor atypical. Such features of our universe then simply cannot be explained by an appeal to typicality and require some other e.g. causal explanation. 

In CM, the natural typicality measure is the Lebesgue measure, respectively the induced microcanonical measure on the energy shell. However, one needs to answer the question how this particular choice is justified. What determines a good typicality measure? We take \emph{stationarity} to be the crucial desideratum, since it is essential to a sensible notation of typicality that it does not change with time. Stationarity of the measure, i.e. equation \eqref{stationarity}, assures that typical sets remain typical and atypical sets remain atypical under the time evolution. In Hamiltonian mechanics, the Lebesgue measure is thus distinguished as the simplest stationary measure on classical phase space.

In the literature, the choice of the Lebesgue measure -- i.e. the equidistribution -- as typicality measure is often motivated by an appeal to the \emph{principle of indifference} (see e.g. \cite{Bricmont}). We do not endorse this point of view. For us, stationarity is the key requirement. When we come to Bohmian QM, we will see that it is indeed the stationary measure, not the equidistribution, that yields the correct notion of typicality.

Historically, the microcanonical measure has also been justified by an appeal to the ergodic hypothesis (see e.g. \cite{Sklar} for a discussion). Ergodicity, in the modern mathematical sense, does indeed distinguish a unique stationary measure (up to sets of measure $0$). However, there is much doubt as to whether the systems usually studied in statistical mechanics are actually ergodic. Also, contrary to widespread believe, ergodicity \emph{per se} has no bearing on Boltzmann's statistical arguments. 

The appeal to stationarity is usually good enough to ensure that the typicality measure is not introduced \emph{ad hoc} but suggested by the fundamental physical theory. Stationarity alone will in general not distinguish the measure uniquely. (For instance, any function of the Hamiltonian $H$ defines a stationary measure on classical phase space). However, since typicality is not \emph{per se} part of the physical laws but a way of reasoning about these laws, the appeal to somewhat subjective criteria such as simplicity is quite appropriate. Still, the situation is more satisfying in Bohmian Mechanics, where the pertinent typicality measure can be shown to be unique in a rather strong sense. 

\subsection{Boltzmann's statistical mechanics: the ideal gas}

\noindent To demonstrate how a typicality argument works, let's consider the stock example of an ideal gas in a box (with perfectly reflecting walls) that will serve as our toy-model for the universe. The number of particles in such a macroscopic system is of the order of Avogadro's constant, that is $N \sim 10^{23}$. Clearly, determining the actual configuration and / or predicting the trajectories for so many particles is a hopeless task, even if the particles are non-interacting as in our example. 

Nevertheless, it is possible to make meaningful predictions about this system. For instance, we can ask the following: what is the rate of particles that have a velocity in $x$-direction that is approximately $v_0$, where $v_0$ is some arbitrary, positive number? We can formalize this in terms of the random variable:

\begin{equation} F(X) := \frac{1}{N}\sum\limits_{i=1}^N \chi_{\lbrace v_{i,x} \in [v_0 - \delta, v_0 + \delta] \rbrace} (X). \end{equation}

\noindent Here, $\delta > 0$ is a small positive number (giving precise meaning to ``approximately $v_0$'') and $\chi$ is the indicator function, i.e. $\chi_{\lbrace v_{i,x} \in [v_0 - \delta, v_0 + \delta] \rbrace}$ equals one if $v_{i,x} = \frac{1}{m} p_{i,x}$ lies in the interval $[v_0 - \delta, v_0 + \delta]$ and zero if it does not. 

Fixing the mean energy per particle to $\frac{E}{N} = \frac{3}{2}k_{\rm B} T$ ($k_{\rm B}$ is the Boltzmann constant and $T$ can later be identified as the temperature of the system), it is a mathematical fact that 
\begin{equation} 
\lim\limits_{N\to\infty, \frac{E}{N}=\frac{3}{2}k_{\rm B} T} \lambda_E \Bigl(\Bigl\lbrace X \in \Gamma_E : v_{i,x} \in [a,b] \Bigr\rbrace \Bigr)
=  \int_a^b
 \frac{\exp \left( - \frac{1}{k_{\rm B} T} \frac{m v^2}{2} 
	\right)}{\Bigl(\frac {2 \pi k_{\rm B} T}{m} \Bigr)^{3/2}}  \mathrm{d} v \,.
	\end{equation}

\noindent From this, one can conclude that for any $\epsilon >0$:

\begin{equation}\label{Maxwellapprox} \lambda_E \Bigl(\Bigl\lbrace X :  \Bigl\lvert \frac{1}{N} \sum\limits_{i=1}^N \chi_{\lbrace v_{i,x} \in [a,b]\rbrace}(X) -  \int_a^b
\frac{\exp \left( - \frac{1}{k_{\rm B} T} \frac{m v^2}{2} 
	\right)}{\Bigl(\frac {2 \pi k_{\rm B} T}{m} \Bigr)^{3/2}}  \mathrm{d} v \Bigr \rvert > \epsilon \Bigr\rbrace\Bigr) \to 0, \, N \to \infty. \end{equation} 

\noindent The derivation of this result is a more or less elementary exercise in measure theory. The more profound question, however, is what this result actually means. 

The function $\rho(v) \propto \exp \left( - \frac{1}{k_{\rm B} T} \frac{m v^2}{2} \right)$ is called the \emph{Maxwell distribution}. It is a probability measure, describing a distribution of particle velocities. Note that there is actually nothing random about the velocities of particles in a gas. The velocity (as well as the position) of every single particle is comprised in the microstate $X$ whose evolution is described by a deterministic equation of motion. There are possible $X$ for which the actual distribution of velocities in the gas differs significantly from that described by the Maxwell distribution. For instance, there are microstates $X$ for which all particles move with one and the same velocity. Or microstates $X$ for which a few very fast particles account for almost the entire kinetic energy, while all the others are nearly at rest. But these states are (obviously) very special ones. The crucial and remarkable fact expressed by equation \eqref{Maxwellapprox} is that, for large $N,$ the \emph{overwhelming majority} of possible microstates is such that the distribution of velocities in the gas is (approximately) Maxwellian. The ``overwhelming majority of microstates'' is thereby defined in terms of the stationary measure $\lambda_E$. In this sense, the Maxwell distribution constitutes a prediction of the microscopic particle theory as a statistical regularity manifested for typical (initial) configurations. 

Ludwig Boltzmann expressed this reasoning as follows: 

\begin{quotation}The ensuing, most likely state [...] which we call that of the Maxwellian velocity distribution, since it was Maxwell who first found the mathematical expression in a special case, is not an outstanding singular state, opposite to which there are infinitely many more non-Maxwellian velocity-distributions, but it is, to the contrary, distinguished by the fact that by far the largest number of possible states have the characteristic properties of the Maxwellian distribution, and that compared to this number the amount of possible velocity-distributions that deviate significantly from Maxwell's is vanishingly small. \cite[p. 252, translation by the authors]{Boltzmann} \end{quotation}

\noindent Note that the role of the microcanonical measure in this argument is only to give precise meaning to ``by far the largest number of all possible states'', that is, to provide a well-defined notion of \emph{typicality}. The Maxwell distribution, in contrast, refers to actual statistical patterns, that is, relative frequencies in typical particle ensembles. Hence, it is important to appreciate the fact that while two measures appear in the mathematical equation \eqref{Maxwellapprox}, their status is very different (c.f. \cite{Goldstein2012}). To make this point clear, we add the following observations:

\begin{enumerate}
	\item Since the box in our example exists only once  -- even more so if it is supposed to be a model for the universe --, probabilistic statements about its (initial) microstate have no empirical meaning. The Maxwellian $\rho$ refers to an actual distribution of velocities that exists in the box. The microcanonical measure does \emph{not} refer to an ensemble of boxes, but pertains to a way of reasoning about the box and the physical laws describing it, allowing us to establish that the observed velocity distribution is typical. 
	
	\item Also, the microcanonical measure is not supposed to quantify our knowledge and / or ignorance about the microstate of the gas. While it is correct to say, in some sense, that randomness in a deterministic theory is only due to our ignorance regarding initial conditions, it is important to note the very limited degree to which \emph{knowledge}, \emph{information}, \emph{credences} or other subjective notions play a role in the analysis. It is an objective fact that for the great majority of microstates, the distribution of velocities in an ideal gas is (approximately) Maxwellian and it is this objective fact that we take to be explanatory.

	\item With respect to a typicality measure, only sets of very large ($\approx 1$) or very small ($\approx 0$) measure are meaningful. Therefore, a probability measure has actually too much mathematical structure and the meaning of ``typical'' would not change, if we changed our measure in a more or less continuous fashion.
\end{enumerate}

\subsection{The coin toss}

An analogous reasoning can be applied to more mundane examples like the before mentioned coin toss. It is a statistical regularity found in our universe that the relative frequency of heads or tails in a long series of fair coin tosses is approximately $1/2$. Now if we agree that a coin toss is guided by the same laws as all other physical processes in the world, this statistical regularity has to be explained on the basis of the fundamental microscopic theory (here: classical mechanics). It is not a new kind of law that holds over and above the microscopic laws. 

Let's denote by $F_i$ the outcome of the i'th coin toss in a long series of $N$ coin tosses. We say that $F_i = 1$ if the outcome is heads and $F_i = 0$ if the outcome is tails. Since classical mechanics is deterministic, the outcome of every single coin toss is actually determined, through the fundamental laws of motion, by the initial state of the universe. Hence, we have: $F_i = F_i(X)$ for $X \in \Gamma$ the initial microstate of the Newtonian universe. The functions $F_i$ are obviously (very) coarse-graining. We do not care about the exact configuration of atoms making up the coin, we do not even care about the exact position or orientation of the coin, we only ask which side is up as the coin lands on the floor. This defines our macroscopic observables. 

There are possible initial configurations conceivable that would give rise to a universe that looks pretty much like ours but in which the relative frequency of heads is very different from $1/2$. Conceivably, there are possible initial configurations for which \emph{every} coin ever to be tossed will land on heads, or for which tails will come out 2 out of 3 times and so on and so forth. But such initial conditions are very special ones. In contrast, \emph{typical} initial conditions of the universe -- compatible with there being coins and coin tossers in the first place\footnote{Actually, in a typical Newtonian universe, there are no coins in the first place, because such universes are in thermal equilibrium. Hence, we would really have to condition our measure on the \emph{past hypothesis}, the low-entropy initial macrostate of the universe, see \cite{Albert:2003}. }-- are such that the relative frequency of heads or tails in a long series of fair coin tosses is approximately $1/2$. Formally, the claim is that for any $\epsilon > 0$,
\begin{equation}\label{cointoss} \lambda \Bigl(\Bigl \lvert \frac{1}{N} \sum\limits_{i=1}^N F_i(X) - \frac{1}{2} \Bigr \rvert > \epsilon \Bigr) \to 0 , \; N \to \infty.
\end{equation} 
This is to say that if $N$ is sufficiently large, the set of initial conditions for which the relative frequency of heads deviates significantly from $1/2$ is extremely small. Such initial conditions are thus not \emph{impossible}, but \emph{atypical}. 

The mathematically trained reader will certainly identify \eqref{cointoss} as a law of large numbers statement. The law of large number is what connects probabilities to relative frequencies in typical ensembles. The distinction between the typicality measure and the probability distribution is here, once again, crucial in order to avoid the usual redundancy of explaining probabilities in terms of probabilities.

We emphasize that, according to this account, \emph{probabilities are objective}. They apply to patterns in the world instead of subjective beliefs. It is a matter of fact that, as the number of coin tosses $N$ becomes very large, \emph{almost all} sequences of coin toss outcomes manifest the pattern of an approximately equal frequency of heads and tails. This matter of fact is independent of what agents believe about the outcomes (although both are linked: it is of course rational to adapt one's beliefs to the patterns in the world). 

Finally, one may wonder why we have described the outcomes $F_i$ as random variables on the configuration space of the \emph{entire universe}. While this may seem a little excessive at first, it is actually where any consistent analysis leads us if one thinks it through to the end. To avoid any queries regarding free will, let us assume that the coins are not tossed by human hand but by a coin-tossing machine. At time $t=0$ a large number $N$ of fair coins is filled into the machine, which is then sealed and shielded from outside influences. From there on, everything takes its (deterministic) course: the outcome of each coin toss is completely determined by the initial configuration of the machine. But the initial configuration of the coin-tossing machine is itself the result of physical processes (the processes of building and setting up the machine) that are determined by suitably specified initial conditions. And these initial conditions are the result of other deterministic processes in an even larger system -- and so on and so forth. To defer the question of typicality further and further to larger and larger systems is just to pass the buck. But the buck stops with the universe. The universe is what it is. There is nothing before and nothing outside. Hence, the key question -- in fact, the only important question -- is whether the statistical patterns we observe are a feature of typical (Newtonian) universes. 

\subsection{Deterministic subsystems: the stone throw}

As mentioned before, there are many situations in CM that are not like the coin toss or the molecules in a gas. For example, when we compute the trajectory of a stone thrown on earth, we can, in general, use a simple deterministic equation without being embarrassed by our ignorance regarding the exact initial microstate of the stone or its environment. There are two conditions satisfied here that allow us to do that:

\begin{enumerate}

\item The external forces, that is, the influence of the rest of the universe neglected in the computations is very small compared to the attraction between the stone and the earth. This is usually the case because other gravitating bodies are either very far away or have very small mass compared to the earth. Formally, this is to say that
\begin{equation}\label{VextCM} V_{ext} \approx 0, \end{equation} 
\noindent which allows us to treat the system stone / earth for all practical purposes as an independent Newtonian intertial system.

\item The evolution of the relevant macroscopic variable -- here, the center of mass of the stone -- is reasonably robust against variations in the microscopic initial conditions. In other words, small changes in the microscopic initial conditions have only small effects on the trajectory of the stone. This is why our ignorance about the exact position and momentum of every single particle constituting the stone (or the earth, or the person/apparatus throwing the stone) does not prevent us from making pretty reliable predictions about the motion of its centre of mass. 
\end{enumerate}

\noindent Nonetheless, even in this case, our prediction for the trajectory of the stone is strictly speaking a typicality result. Atypical events in the environment or fluctuations of the particles constituting the stone could lead to very different outcomes. Hence, to be precise, we would have to cast our result about the trajectory of the stone in a form that looks quite similar to the probabilistic statements \eqref{Maxwellapprox} or \eqref{cointoss}. For instance, denoting by $x(t)$ the computed trajectory (depending on the initial position and momentum of the stone) and by $\tilde x(t)$ the actual trajectory of the stone (depending on the initial condition $X$ of the universe), we could write:

\begin{equation} \lambda \Bigl(\Bigl\lbrace X : \sup\limits_{0 \leq t \leq T} \lvert \tilde x(t) - x(t) \rvert > \epsilon \Bigr\rbrace \Bigl) \approx 0. \end{equation}

\noindent Still, the stone throw example points to a striking difference between classical and quantum mechanics. In CM, we often encounter situations in which correlations between a subsystem and its environment become negligible, allowing for a (more or less) deterministic description of the subsystem. In QM, by contrast, the generic situation is much more similar to the coin toss or the molecule in a gas, where predictions of statistical patterns are the best we can hope for. Our aim is now to explain why this is so. 

\section{Probabilities in Bohmian quantum mechanics}
\label{sec:prob-BM}

Having discussed probabilities in classical physics, we now turn to the quantum case. In QM, we encounter a new dynamical feature that is totally absent from CM: the specification of initial positions and momenta is replaced with the specification of an initial wave function. The wave function is entangled and is defined on configuration space. Due to entanglement, wave functions cannot be attributed to the particles separately, as initial parameters are attributed to them separately in CM. That is to say, there is fundamentally only one wave function for the whole particle configuration of the universe taken together. It correlates the state of any particle with, in principle, any other particle, without that correlation having to depend on the distance between the particles.  
 
As announced in the introduction, we consider Bohmian Mechanics (BM) to provide the most convincing solution to the quantum measurement problem. BM is based on the following three axioms:
  
\begin{enumerate}
    \item A Bohmian system with $N$ particles is completely described by a couple $(Q,\Psi)$, where $Q=(Q_1, \dots, Q_N)\in\mathbb R^{3N}$ represents the spatial configuration of the particles and $\Psi$ is a complex, square-integrable function on the configuration space $\IR^{3N}$ called the universal wave function. 
    
    \item The evolution of the wave function $\Psi$ is described by the Schrödinger equation
    \begin{align}
    \label{eq:schroedinger}
    i\hbar \partial_t \Psi_t
    =
    H\Psi_t,  
    \end{align}
  where $H$ is the Hamiltonian of the system. 
  
    \item The evolution of the particle configuration is described by a first order differential equation in which the wave function $\Psi$ enters to determine a velocity field $v^{\Psi}_t$ along which the particles move. More precisely, the particles move according to the \emph{guiding equation}
    \begin{align}
    \label{eq:BM-vel}
    \dot{Q}_k = v^\Psi_{k,t}(Q) := 
    \frac{\hbar}{m_k}\Im \frac{\nabla_k\Psi_t(Q)}{\Psi_t(Q)},
    \end{align}
    
   where $m_k$ denotes the mass of the $k$'th particle. Note that, due to the entanglement of the wave function, the resulting law of motion is non-local. 
\end{enumerate}

\noindent Given an initial wave function $\Psi_0$ and the initial particle configuration $Q_0\in\mathbb R^{3N}$, the evolution of the system is completely and uniquely determined for all times. This determinism is contrary to the popular believe that quantum mechanics is intrinsically and irreducibly random. However, since we do not know (in fact, as we will see, \emph{cannot} know) the exact particle configuration, we have to resort once again to a statistical analysis in order to extract meaningful predictions. We will now show that, to this end, we can pursue the same strategy as we did before in CM. In the following, we will largely rely on the development of this strategy in \cite{Durr:2009fk} and \cite{Durr:2013aa} (see \cite{Callender:2007aa} and \cite{Maudlin:2007ab} for a philosophical analysis). 

For a statistical analysis of BM, we need a) a sensible typicality measure defined on configuration space and b) a procedure to get from the fundamental, universal description in terms of the universal wave function to a well-defined description of Bohmian subsystems. Given the universal wave function, the appropriate notion of typicality for particle configurations is given in terms of the measure with density $\left|\Psi\right|^{2}$. The crucial feature of this measure is that it is \emph{equivariant}, assuring that typical sets remain typical and atypical sets remain atypical under the Bohmian time-evolution. More precisely, if $\Phi^\Psi_{t,0}$ is the flow on configuration space induced by the guiding equation \eqref{eq:BM-vel}, then

\begin{equation}\mathbb{P}^{\Psi}(A):= \int\limits_{A} \left|\Psi_0\right|^{2} \mathrm{d}^{3N} q = \int\limits_{\Phi^\Psi_{t,0}(A)}  \left|\Psi_t\right|^{2} \mathrm{d}^{3N} q \end{equation}
holds for any measurable set $A \subseteq \IR^{3N}$. Equivariance is thus the natural generalization of stationarity for a non-autonomous (time-dependent) dynamics. The $\left|\Psi\right|^{2}$-measure can be proven to be the unique equivariant measure for the Bohmian particles dynamics that depends only locally on $\Psi$ or its derivatives (see \cite{Goldstein2007}). In this sense, it is even more strongly suggested as the correct typicality measure for BM than the Lebesgue measure was in CM. 

Let us now have a closer look at how BM treats subsystems of the universe. Suppose that the subsystem consists of $n \ll N$ particles. We then split the configuration space into $\IR^{3N}= \IR^{3n} \times \IR^{3(N-n)}$, so that, writing $q=(x,y)$, the $x$-coordinates describe the degrees of freedom of the subsystem and the $y$-coordinates describe the possible configurations of the rest of the universe. Analogously, we split the \emph{actual} particle configuration into $Q=(Q^{sys},Q^{env})=(X,Y)$, with $Q^{sys}=X,$ the configuration of the subsystem under investigation and $Q^{env}=Y$ the configuration of its environment. 

Now, in passing from the fundamental, universal theory to a description of the subsystem, we can just take the universal wave function $\Psi_t(q) = \Psi_t(x,y)$ and plug into the $y$ argument the actual configuration $Y(t)$ of the rest of the universe. The resulting 
\begin{equation}\psi^Y_t(x):=\Psi_t(x,Y(t)) \end{equation} 
is now a function of the $x$ coordinates only. It is called the \emph{conditional wave function}. In terms of this conditional wave function, the equation of motion for the subsystem takes the form
\begin{align}
\label{eq:X-vel}
 \dot{X}(t) \propto \Im \frac{\nabla_x\psi^Y_t(x)}{\psi^Y_t(x)}\Biggl\lvert_{x=X(t)} 
\end{align} 

\noindent to be compared with \eqref{eq:BM-vel}. However, since the conditional wave function depends explicitely on $Y(t)$, its time-evolution may be extremely complicated and not follow any Schrödinger-like equation. Fortunately, in many relavant situations, the subsystem will dynamically decouple from its environment. We say that the subsystem has an \emph{effective wave function} $\varphi$ if the universal wave function takes the form 
\begin{align}\label{effectivewf}
\Psi(x,y)=\varphi(x)\chi(y)+\Psi^{\perp}(x,y),
\end{align}
where $\chi $ and $\Psi^{\perp}$ have disjoint $y$-support and $Y \in \mathrm{supp} \, \chi$, so that in particular $\Psi^{\perp}(x,Y) = 0$ for almost all $x$. (Note that this is much weaker than assuming that $\Psi$ has a product structure, which is in general not the case.) This means that we can effectively forget about the empty wave packet $\psi^\perp{(x,y)}$ and describe the subsystem in terms of its own independent wave function $\varphi$. If we can furthermore assume that the interaction between subsystem and environment is negligible for some time, that is
\begin{equation}\label{VextQM} V_{ext}(x,y)\varphi(x)\chi(y) \approx 0, \end{equation}

\noindent the effective wave function will satisfy its own, autonomous Schrödinger evolution.\footnote{From the point of view of the subsystem, this part of the interaction potential coupling $x$ and $y$ degrees of freedom is the external potential. Condition \eqref{VextQM} is thus the same as \eqref{VextCM} above.} Such a $\varphi$ -- normalized to $\int \lvert \varphi(x) \rvert^2 \mathrm{d} x = 1$ -- is the Bohmian counterpart of the usual quantum mechanical wave function. It is these effective wave functions that physicists manipulate in laboratories and for which Born's rule is formulated.

For our statistical analysis, we start by considering the conditional measure
\begin{align}\label{PPsi}
\mathbb{P}^\Psi(\{{Q}=({X},{Y}),{X}\in
\mathrm{d}^n x\}|{Y}) = \frac{|\Psi(({x},{Y}))|^2\mathrm{d}^n x}{\int |\Psi(({x},{Y}))|^2 \mathrm{d}^n x} = |\psi^{Y}({x})|^2 \mathrm{d}^nx.
\end{align}
In the special situations described by \eqref{effectivewf}, the conditional wave function $\psi^{Y}$ on the right hand side becomes the effective wave function $\varphi$. This formula already holds a very deep inside to which we will return in a while. For practical purposes, though, conditioning on the configuration $Y$ is much too specific, since we have only very limited knowledge of $Y$. However, many different $Y$ will yield one and the same effective wave functions for the subsystem. Collecting all those $Y$, and using the fact that by yielding the same effective wave function they also yield the same conditional measure \eqref{PPsi}, a simple identity for conditional probabilities yields
\begin{equation}  \label{PPsi''}
\mathbb{P}^\Psi(\{{Q}=({X},{Y}),{X}\in
\mathrm{d}^nx\}|\psi^{ Y}=\varphi)= |\varphi|^2
\mathrm{d}^nx.   \,\,
\end{equation} 

\noindent From this formula, one can now derive law of large numbers estimates of the following kind: at a given time $t$, consider an ensemble of $M$ identically prepared subsystems with effective wave function $\varphi$. Denote by $X_i$ the actual configuration of the i'th subsystem. Let $A \subseteq \IR^{3n}$ consider the corresponding indicator function $\chi_{\lbrace X_i \in A \rbrace}$, which is $1$, if the configuration $X_i$ is in $A$ and $0$ otherwise. Then it holds for any $\epsilon >0$ that

\begin{align}
\mathbb{P}_t^\Psi=\Bigl(\Bigl\lbrace Q : \Bigl \lvert \frac{1}{N}\sum^{N}_{i=1} \chi_{\lbrace X_i \in A \rbrace}(Q)-\int_{A} |\varphi(x)|^2 \Bigr\rvert <\epsilon \Bigr\rbrace\Bigr) \to 0, N \to \infty.
\end{align}

\noindent This is to say that for \emph{typical} configurations of the universe, the particles in an ensemble of subsystems with effective wave function $\varphi$ are distributed according to $\lvert \varphi \rvert^2$. Thus, Born's rule holds in typical Bohmian universes, that is, in \emph{quantum equilibrium}.

Once again we emphasize that the $\lvert \Psi \rvert^2$-measure given in terms of the \emph{universal} wave function is only used to define typicality. It is \emph{not} supposed to describe an actual distribution of configurations, that is, an ensemble of universes, because the universe exists only once. By contrast, the $\lvert \varphi \rvert^2$-measure on the right hand side, defined in terms of the effective wave function, does refer to actual particle distributions in a typical ensemble of identically prepared subsystems. Born's rule is thus predicted and explained by BM as a statistical regularity of typical Bohmian universes.

Comparing equation \eqref{PPsi''} to \eqref{Maxwellapprox} (and recalling the reasoning that lead to the respective equations) we recognize the analogy between the derivation of Maxwell's distribution in CM and Born's rule in Bohmian QM. In essence, it is Boltzmann's statistical mechanics applied to two different theories. The status of probabilities and the role of typicality is the same in both cases, although the dynamical laws are strikingly different. On the one hand, this illustrates the deepness and universality of Boltzmann's insights. On the other hand, it shows that there is no need to look for a fundamentally new kind of randomness in the quantum realm. If the microscopic laws and the ontology of the theory are clear, probabilities in QM are no more mysterious than they are in CM. 

\section{Conclusion}
\label{sec:conclusion}

So far, we have highlighted the similarities between the statistical analysis of CM and Bohmian QM, showing that probabilities have the same status in both theories. But what then is the difference between classical mechanics and quantum mechanics? Why is it that the quantum realm appears to us so much more random and unpredictable?

The answer to this question is in part trivial. QM is usually employed to make predictions about microscopic systems, while CM is most often employed to make predictions about macroscopic systems and coarse-grained observables. The latter are bound to be more robust against our ignorance regarding microscopic initial conditions. Furthermore, our ability to describe a particular subsystem and the level of detail that we can thereby achieve depends heavily on the strength of correlations between the investigated subsystem and the rest of the universe. Newtonian mechanics is a non-local theory, though only in a rather mild sense. Forces fall off quickly with increasing distance (and gravity is very weak to begin with) so that parts of the universe can often be described as autonomous Newtonian systems for all practical purposes.

In quantum mechanics, by contrast, non-locality is much more prevalent. This is clearly brought out by BM, where the configuration of particles is guided by a common wave function so that the velocity of any particle depends, in general, on the position of all the other particles. In any case, due to entanglement, QM allows for correlations that do not depend on the distance between the correlated systems (the best known example being the spin singlett state, leading to the famous anti-coincidences in the EPRB experiment). This makes it much more difficult to consider any proper part of a Bohmian universe as ``isolated'', while ignoring the influence of the rest of the universe.  As a matter of fact, it is often possible to provide an autonomous Bohmian description of a Bohmian subsystem in terms of an effective wave function. This autonomy, however, can be somehow deceiving, because the effective wave function still depends implicitly on the configuration of the environment (e.g. on the procedure used to prepare that state in an experimental situation). 

More precisely (and more profoundly), our possible knowledge about the particle configuration in a Bohmian subsystem is restricted by the theorem of \emph{absolute uncertainty}, which has no analog in classical physics (see \cite{Durr:2013aa}, chapter 2). Absolute uncertainty is a direct consequence of the conditional probability formula \eqref{PPsi}: all our \emph{records} about the particle positions -- brain states, computer prints, pointer position, etc. -- are included in the configuration $Y$ of the rest of the universe. Hence, all possible correlations between these records and the configuration of the subsystem are already taken into account in equations \eqref{PPsi} and \eqref{PPsi''} that yield Born's rule for the distribution of particle positions. 

This connection between our epistemic state and the effective wave function of the subsystem then works in two ways. One the one hand, it means that given a Bohmian subsystem with effective wave function $\varphi$, our information about the particle configuration cannot be more precise than what is given by the $\lvert \varphi \rvert^2$-distribution. On the other hand, it means that if we perform additional measurements to determine the particle positions with greater accuracy, the system's effective wave function becomes more and more peaked. Hence, the gradients in the velocity formula \eqref{eq:BM-vel} induce higher and higher possible velocities, depending on the precise initial configuration of the particles. Less uncertainty about the initial particle positions thus implies more uncertainty about the (asymptotic) velocities -- this is the source of Heisenberg's uncertainty principle. Even small deviations in the initial configuration will thus lead to large deviations of the resulting Bohmian trajectories.\footnote{Our rapidly increasing uncertainty about the particle positions is then mirrored by the quick spreading of the wave function under the Schrödinger time evolution.} In other words, the manifestly non-local nature of quantum mechanics is such that a system becomes immediately more chaotic as we try to minimize our ignorance regarding microscopic initial conditions. As a consequence, we have to resort to probabilistic reasoning much earlier than is often the case in classical physics. For a quantum system, Born's rule provides -- provably -- as good a description as we can get in a universe in quantum equilibrium.

\paragraph{Acknowledgements.} We would like to thank an anonymous referee for valuable comments on the first version of this paper. We are grateful to Detlef Dürr, Sheldon Goldstein and Nino Zanghì for helpful discussions on the topic of this paper. Andrea Oldofredi's work was supported by the Swiss National Science Foundation, grant no. 105212-149650, Dustin Lazarovici's work was supported by the Cogito Foundation, grant no. 15-106-R, while D.-A.\ Deckert's work was funded by the junior research group grant \emph{Interaction between Light and Matter} of the Elite Network of Bavaria.

\bibliographystyle{apalike}
\bibliography{references_fundont}

\begin{thebibliography}{}

\bibitem[Albert, 2003]{Albert:2003}
Albert, D.~Z. (2003).
\newblock {\em Time and Chance}.
\newblock Cambridge, Massachusetts: Harvard University Press.

\bibitem[Bohm, 1952]{Bohm:1952aa}
Bohm, D. (1952).
\newblock A suggested interpretation of the quantum theory in terms of
  ``hidden'' variables. 1.
\newblock {\em Physical Review}, 85(2):166--179.

\bibitem[Boltzmann, 1896]{Boltzmann}
Boltzmann, L. (1896).
\newblock {\em Vorlesungen \"uber Gastheorie}.
\newblock Verlag v. J. A. Barth, Leipzig.

\bibitem[Bricmont, 1995]{Bricmont}
Bricmont, J. (1995).
\newblock Science of chaos or chaos in science?
\newblock {\em Annals of the New York Academy of Sciences}, 775(1):131--175.

\bibitem[Callender, 2007]{Callender:2007aa}
Callender, C. (2007).
\newblock The emergence and interpretation of probability in bohmian mechanics.
\newblock {\em Studies in History and Philosophy of Modern Physics},
  38:351--370.

\bibitem[de~Broglie, 1928]{Broglie:1928aa}
de~Broglie, L. (1928).
\newblock La nouvelle dynamique des quanta.
\newblock {\em Electrons et photons. Rapports et discussions du cinqui{\`e}me
  Conseil de physique tenu {\`a} Bruxelles du 24 au 29 octobre 1927 sous les
  auspices de l'Institut international de physique Solvay}, pages 105--132.
\newblock Paris: Gauthier-Villars. English translation in Bacciagaluppi, G. and
  Valentini, A., editors (2009). \emph{Quantum theory at the crossroads.
  Reconsidering the 1927 Solvay conference}, pages 341--371. Cambridge:
  Cambridge University Press.

\bibitem[D{\"u}rr et~al., 2013]{Durr:2013aa}
D{\"u}rr, D., Goldstein, S., and Zangh{\`\i}, N. (2013).
\newblock {\em Quantum physics without quantum philosophy}.
\newblock Berlin: Springer.

\bibitem[D{\"u}rr and Teufel, 2009]{Durr:2009fk}
D{\"u}rr, D. and Teufel, S. (2009).
\newblock {\em Bohmian mechanics: the physics and mathematics of quantum
  theory}.
\newblock Berlin: Springer.

\bibitem[Esfeld, 2014]{Esfeld:2014ac}
Esfeld, M. (2014).
\newblock The primitive ontology of quantum physics: guidelines for an
  assessment of the proposals.
\newblock {\em Studies in History and Philosophy of Modern Physics},
  47:99--106.

\bibitem[Goldstein, 2012]{Goldstein2012}
Goldstein, S. (2012).
\newblock Typicality and notions of probability in physics.
\newblock In Ben-Menahem, Y. and Hemmo, M., editors, {\em Probability in
  Physics}, pages 59--71. Springer Berlin Heidelberg.

\bibitem[Goldstein and Struyve, 2007]{Goldstein2007}
Goldstein, S. and Struyve, W. (2007).
\newblock On the uniqueness of quantum equilibrium in bohmian mechanics.
\newblock {\em Journal of Statistical Physics}, 128(5):1197--1209.

\bibitem[Lazarovici and Reichert, 2015]{Typicality}
Lazarovici, D. and Reichert, P. (2015).
\newblock Typicality, irreversibility and the status of macroscopic laws.
\newblock {\em Erkenntnis}, 80(4):689--716.

\bibitem[Maudlin, 1995a]{Maudlin:1995aa}
Maudlin, T. (1995a).
\newblock Three measurement problems.
\newblock {\em Topoi}, 14:7--15.

\bibitem[Maudlin, 1995b]{Maudlin:1995ab}
Maudlin, T. (1995b).
\newblock Why {B}ohm's theory solves the measurement problem.
\newblock {\em Philosophy of Science}, 62:479--483.

\bibitem[Maudlin, 2007]{Maudlin:2007ab}
Maudlin, T. (2007).
\newblock What could be objective about probabilities?
\newblock {\em Studies in History and Philosophy of Modern Physics},
  38:275--291.

\bibitem[Schilpp, 1970]{schilpp}
Schilpp, P., editor (1970).
\newblock {\em Albert {E}instein: {P}hilosopher-{S}cientist}.
\newblock Number~1 in The Library of living philosophers. Open Court Press, 3rd
  edition.

\bibitem[Sklar, 1973]{Sklar}
Sklar, L. (1973).
\newblock Statistical explanation and ergodic theory.
\newblock {\em Philosophy of Science}, 40(2):194--212.

\end{thebibliography}

\end{document}